# QUERIES MINING FOR EFFICIENT ROUTING IN P2P COMMUNITIES

Anis ISMAIL[1], Mohamed QUAFAFOU[1], Nicolas DURAND[1], Gilles NACHOUKI[2] and Mohammad HAJJAR[3]

[1]LSIS, Domaine Universitaire de Saint-Jérôme, Marseille, France
`{anis.ismail, mohamed.quafafou,`
`nicolas.durand}@univmed.fr`
[2] LINA Laboratory, Nantes, France
`gilles.Nachouki@univnantes.Fr`
[3] Lebanese University, Lebanon
`m_hajjar@ul.edu.lb`

## ABSTRACT

*Peer-to-peer (P2P) computing is currently attracting enormous attention. In P2P systems a very large number of autonomous computing nodes (the peers) pool together their resources and rely on each other for data and services. Peer-to-peer (P2P) Data-sharing systems now generate a significant portion of Internet traffic. Examples include P2P systems for network storage, web caching, searching and indexing of relevant documents and distributed network-threat analysis. Requirements for widely distributed information systems supporting virtual organizations have given rise to a new category of P2P systems called schema-based. In such systems each peer exposes its own schema and the main objective is the efficient search across the P2P network by processing each incoming query without overly consuming bandwidth. The usability of these systems depends on effective techniques to find and retrieve data; however, efficient and effective routing of content-based queries is a challenging problem in P2P networks. This work was attended as an attempt to motivate the use of mining algorithms and hypergraphs context to develop two different methods that improve significantly the efficiency of P2P communications. The proposed query routing methods direct the query to a set of relevant peers in such way as to avoid network traffic and bandwidth consumption. We compare the performance of the two proposed methods with the baseline one and our experimental results prove that our proposed methods generate impressive levels of performance and scalability.*

## KEYWORDS

*Peer-to-peer systems, Data mining, hypergraphs, Query routing*

## 1. INTRODUCTION

The traditional P2P systems [12] [30] [22] [8] offer support for richer queries than just search by identifier, such as keyword search with regular expressions. In recent years, P2P has emerged as a popular way to share huge volumes of data [3], [41]. The major problem in such networks is query routing, i.e. deciding to which other (Super-)Peers the query has to be sent for high efficiency and effectiveness [29]. However, systems that broadcast all queries to all Peers suffer from limited efficiency and scalability.

The purpose of a data-sharing P2P system is to accept queries from users, locate, and return data (or pointers to the data) to the users. Each Peer owns data (expertise) to be shared with other Peers. The shared data usually consists of files, but is not restricted to files; it could be stored records in a relational database. Queries may take any form that is appropriate given the type of shared data. If the system is a file-sharing system, queries may be file identifiers, or keywords with regular expressions. Nodes, like Super-Peers, process queries and produces results groups of Peers, and the result set for a query is the union of results from every Super-Peer (SP), groups of Peers and their Super-Peers, that processes the query. When a Peer submits a query, this Peer becomes the source of this query. The query is transmitted to its Super-Peer. The routing policy in use determines relevant neighbours (SP) quickly, based on semantic mappings between schemas of (Super-)Peers, and then send the query to them. When a SP receives a query, it will process it over its local collection of data sources considering its different Peers. If at least one of its Peers answers the question then results are found and the SP will send a single response message back to the query source. Another important aspect of the user experience is how long the user must wait for results to arrive. This is due to large part to the mediation process which remains difficult to realize in such a context when the number of (Super-)Peer increases. Response times tend to be slow in hybrid P2P networks, since the query travel through several SP in the network and whenever the SP is forced to look for connections (i.e. mappings) in order to route the query. Satisfaction time is simply the time that has elapsed from when the query is first submitted by the user, to when the user receives the overall results. For a deep discussion of this problem we refer the reader to [4] [3].

This work was intended as an attempt to motivate the use of mining algorithms and hypergraphs context to construct efficient solutions to this query routing problem. Firstly, we have developed a decision tree based method by mining queries and constructing a predictive model for each Super-Peer. As for prerequisites, the reader is expected to be familiar with decision tree based methods [28] [39]. Furthermore, we have constructed clusters of Super-Peers and defined a hypergraphs space that we have used to explicit the minimal transversal where each one contains a set of Super-Peers. The minimally notion is explained formally in the section 2.4 and applied in the P2P context in 4.3. Our main goal is to reduce the processing of queries at the SP level to predict others relevant SP able to process such queries. For this reason, our proposed methods focus on how the query is routed to relevant Peers with minimum query processing in order to improve the answering time.

The following section recalls briefly the problem overview. Section 3 presents the basic notions and concepts as queries routing and data mining in the P2P context, soft-clustering and Hypergraph Transversals. Section 6 presents an overview on Information retrieval in P2P context. The proposed methods for queries routing are introduced in the section 5. Finally the section 6 is dedicated to the discussion of the experimental results and we conclude in the section 7.

## 2. BACKGROUND

To facilitate access to the individual topics, the following sections are rendered as self-contained as possible.

### 2.1. Basic notions

A Peer is an autonomous entity with a capacity of storage and data processing. In a computer network, a Peer may act as a client or as a server. A P2P is a set of autonomous and self-organized Peers (P), connected together through a computer network. The purpose of a P2P network is the sharing of resources (files, databases) distributed on Peers by avoiding the appearance of a Peer as a central server in this network. We note: P2P = (P, U), P is the set of Peers and U represents links (overlay connections) between two Peers $P_i$ and $P_j$, $U \subseteq P \times P$. The hybrid P2P (P2Ph) (See Figure 1) network that we consider in this paper includes sets of Peers (P) and Super-Peers (SP). We note : P2Ph = (P $\cup$ SP, K), where P is the set of Peers, SP is the set of Super-Peers and K is the set of overlay links expressed under the format of pairs : ($P_i$, $SP_j$) or ($SP_j$, $SP_k$) which respectively link a Peer $P_i$ to a Super-Peer $SP_j$ or a Super-Peer $SP_j$ to one or several Super-Peers $SP_k$.

A PDMS (Peer Data Management System) combines P2P systems and databases systems. Each Peer is supposed to hold a database (or an XML document, etc.) with a data schema. Each Super-Peer provides a theme (a semantic domain, a subject, or an idea) representing special interest to a group of Peers. The themes are not necessarily separated; they are described by Super-Peers, with the three following manufacturers:

– A concept is a collection of individuals that constitute the entities of the modelled domain. The concepts can be compared to the notion of class (i.e. object model) or type of entity in the conceptual models (i.e. Entity/Relationship).

– A role is a binary relationship between concepts. Roles are used to specify properties of instances and are compared to the notion of attributes in the conceptual models. A role is viewed as a function linking a concept (called domain) to another concept (known as co-domain).

– Specialization (IsA) starts from a specific concept to a more general concept. It is transitive and asymmetric and defines a hierarchy between concepts it connects.

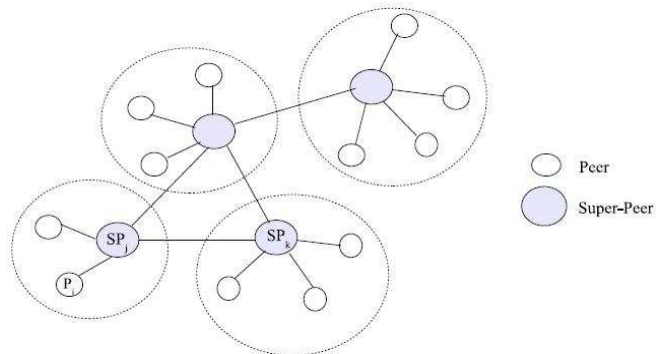

Figure 1. Super-Peer based network.

We note R the set of relations reduced in this paper to two relations that are {Role; IsA} and PDMS={PS $\cup$ SP$_{T, D}$ , K} where PS represents all the Peers of the network with their data schemas S = {S$_1$, …., S$_p$}. A Peer is connected to the network with only one data schema. K is the set of overlay links between (Super-)Peers. Each Peer P $\in$ PS is doted of a Data Management System (denoted DMS) able to manage their data. T={T$_1$,…., T$_k$} represents the interest themes published by Super-Peers SP through the network. In our case, each Super-Peer publishes only one theme and Peers expresses that are interested by one or several theme(s) in T. The themes are not disjoints: two Super-Peers can publish the same concepts or roles with distinct structures and/or don't use the same vocabulary.

D={D$_1$, …., D$_k$} describes the themes in the set of T: D$_j$ describes the theme T$_j$ specifying the set of concepts and their relationships.

## 2.2 Data mining in the P2P context

Knowledge discovery and data mining (KDD) from P2P network is a relatively new field with little related literature. P2P data mining has recently emerged as an area of KDD research, specifically focusing on algorithms which are efficient query routing and scalability. For instance, Raahemi et al. [33] present a new approach using data-mining technique, to classify Peer-to-Peer traffic in IP networks by capturing Internet traffic at a main gateway router. Then, they built several models using a combination of various attribute sets for different ratios of P2P to non-P2P traffic in the training data. Using the same technique, Roussopoulos et al. [14] present a heuristic that designers can use to judge how suitable a P2P solution might be for a particular problem. It is based on characteristics of a wide range of P2P systems from the literature, both proposed and deployed. These include budget, resource relevance, trust, rate of system change, and criticality.

Bhaduri et al. [7] propose an alternate solution that works in a completely asynchronous manner in distributed environments and offers low communication overhead, a necessity for scalability. For more details on the distributed mining approach we refer the reader to [20] that offers a scalable and robust distributed algorithm for decision tree induction in large Peer-to-Peer environments.

Content location is a challenging problem in decentralized Peer-to-Peer systems. And query-flooding algorithm in Gnutella system suffers from poor scalability and considerable network overhead. Currently, based on the Small-world pattern in the P2P system, a piggyback algorithm called interest based shortcuts gets a relatively better performance. However, Xi Tong; Dalu Zhang; Zhe Yang [40] believe it could be improved and become even more efficient, and a cluster-based algorithm is put forward.

The main concern of their algorithm is to narrow the search scope in content location. Resource shortcuts are grouped into clusters according to their contents, and resource queries are only searched in related shortcut clusters, so that the search efficiency is guaranteed and the network bandwidth is saved. In their experiment, cluster-based algorithm uses only 40% shortcuts roughly, compared with the former algorithm and the same success rate is achieved.

## 2.3. Soft-Clustering

In the following discussion, we use the most common terms in KDD: each object corresponds to a data record and is called a transaction, and is described by items (for example, attribute-value pairs). For a transaction, an item has a binary value: present (i.e., the transaction has the characteristic depicted by the item) or not. A pattern is a set of items (also called itemset).

ECCLAT (Extraction of Clusters from Concepts LATtice) [16] discovers overlapping clusters. It produces lists of attributes to describe each discovered cluster of objects. The approach used by ECCLAT is quite different from usual clustering techniques. Unlike existing techniques, ECCLAT does not use a global measure of similarity between elements but is based on the discovery and the evaluation of potential clusters coming from the set of frequent closed patterns [32]. Moreover, the number of resulting clusters is not set in advance. A cluster is composed of a pattern and a set of transactions containing this pattern. A pattern is frequent if its frequency is at least the frequency threshold, noted minfr, set by the user. ECCLAT starts from the set of all frequent closed patterns. Indeed, a closed pattern checks an important property for clustering: it gathers a maximal set of items shared by a set of transactions. In other words, this allows capturing the maximum amount of similarity. These two points (the capture of the maximum amount of similarity and the frequency) are the basis of the approach of clusters selection. ECCLAT evaluates and selects the most interesting patterns by using a cluster evaluation measure. All computations and interpretations are detailed in [16]. The cluster evaluation measure is composed of two criteria: homogeneity and concentration. With the homogeneity value, clusters having many items shared by many transactions are favoured (a relevant cluster has to be as homogeneous as possible and should gather "enough" transactions). The concentration measure limits an excessive overlapping of transactions between clusters. Finally, the interestingness of a cluster is defined as the average of its homogeneity and its concentration. ECCLAT uses the interestingness to select clusters and to produce a clustering with a slight overlapping between clusters. The overlapping depends on the value of a parameter, noted M, corresponding to the minimal number of different transactions between two selected clusters. The algorithm performs as follows. The cluster having the highest interestingness is selected. Then as long as there are transactions to classify (i.e., which do not belong to any selected clusters) and some clusters are left, the cluster, having the highest interestingness and containing at least M transactions not classified yet, is selected. The number of clusters is established by the selection process.

## 2.4 Hypergraph Transversal

Hypergraph theory [6] is one of the most important areas of discrete mathematics with significant applications in many fields of computer science in particular data mining [21].

A hypergraph H is a generalized graph defined by a pair (V, E) where V={$v_1$, $v_2$, …, $v_n$} is a set of vertices and E={$e_1$, $e_2$, …, $e_m$} is a set of non-empty subsets of V called hyperedges. While graph edges are pairs of vertices, hyperedges are arbitrary sets of vertices, and can therefore contain an arbitrary number of vertices. Figure 2 presents an example of hypergraph with six vertices ($v_1$, $v_2$, $v_3$, $v_4$, $v_5$, $v_6$) and three hyperedges (e1= {$v_1$, $v_3$, $v_5$ }, $e_2$={ $v_5$, $v_6$}, e3={$v_1$, $v_2$, $v_4$}).

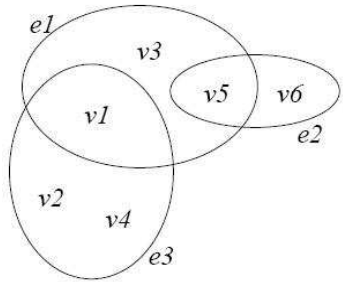

Figure 2. Example of hypergraph.

One of the most problems on hypergraphs is the computation of the transversals. A transversal (or, hitting set) of a hypergraph H = (V, E) is a set T $\subseteq$ V that has non-empty intersection with every hyperedge, i.e., $T \cap e_i \neq \phi$. There is a considerable amount of works on hypergraph transversals, which principally concentrate on the minimal transversals computation [19]. A transversal T is called minimal if no proper subset T' of T is a transversal of H. The set of the minimal transversals of H is noted MinTr(H ). Let us note that (V, MinTr(H )) is a hypergraph called transversal hypergraph [6]. In Figure 2, $v_2v_3v_5$ is a transversal but not a minimal transversal because $v_2v_5$ is a transversal. $v_5$ is common to e1 and $e_2$ and $v_2$ belongs to $e_3$. In our example, MinTr(H) = {$v_1v_5$; $v_1v_6$; $v_2v_5$; $v_4v_5$; $v_2v_3v_6$; $v_3v_4v_6$}.

A minimal transversal can be identified in polynomial time by removing, starting from V, one-by-one the vertices of V and checking after each removal whether the remaining set is a hitting set. However, finding a transversal with minimum cardinality is NP-hard. Indeed, the number of minimal transversals in a hypergraph H can be exponential in |H| = n×m, the size of H. Thus, it does not exist an algorithm computing MinTr(H ) with a polynomial complexity in | H |. Berge [6] is the first to propose an algorithm computing the minimal transversals. This algorithm starts to find the minimal transversals of a hyperedge (i.e., each vertex of the hyperedge), then it adds the other hyperedges one-by-one. After each addition, the minimal transversal set is updated. This algorithm is not practical on large hypergraphs. In the last decade, many algorithms have appeared [5], [15], [24]. They are improvements of the initial algorithm proposed by Berge. A lot of these algorithms use the links (formalized in [21]) between minimal transversals, data mining and machine learning. MTminer [23] is a recent algorithm based on data mining techniques and concept lattices to compute minimal transversals.

A hypergraph is a convenient mathematical structure for modeling numerous problems in both theorical and applied computer science. In [17], hypergraph transversals are used to discover interesting collections of Web services. In [18], hypergraphs model results of data clustering. The vertices represent the clusters and the hyperedges correspond to the clustering results. Minimal transversals are then used to guide a similarity detection process through clustering results.

## 3. Query Routing In P2p Networks

Research in P2P systems, such as Chord [36], CAN [34], Pastry [35] or P-Grid [1] is based on various forms of distributed hash tables (DHTs) and supports mappings from keys, e.g., titles or authors, to locations in a decentralized manner such that routing scales well with the number of Peers in the system.

Lu and Callan [27] Consider content-based retrieval in hybrid P2P networks where a Peer can either be a simple node or a directory node. Directory nodes serve as Super-Peers, which may possibly limit the scalability and self-organization of the overall system. The Peer selection for forwarding queries is based on the Kullback-Leibler divergence between Peer-specific statistical models of term distributions. Strategies for P2P request routing beyond simple key lookups but without considerations on ranked retrieval have been discussed in [44], [11], [10], but are not directly applicable to our setting. The construction of semantic overlay networks is addressed in [27], [11] using clustering and classification techniques; these techniques would be orthogonal to our approach. Tang, Xu, and Dwarkadas [37] distribute a global index onto Peers using LSI dimensions and the CAN distributed hash table. In this approach Peers give up their autonomy and must collaborate for queries whose dimensions are spread across different Peers. [2] addresses the problem of building scalable semantic overlay networks and identifies strategies for their traversal. A good overview of metasearch techniques is given by [38]. [26] discusses specific strategies to determine potentially useful local search engines for a given user query. Notwithstanding the relevance of this prior work, collaborative P2P search is substantially more challenging than metasearch or distributed IR over a small federation of sources, as these approaches mediate only a small and rather static set of underlying engines, as opposed to the high dynamics of a P2P system. Castano and Montanelli addressed the problem of formation of semantic Peer-to-Peer communities [9]. Each Peer is associated with an ontology which gives a semantically rich representation of the interests that the Peer exposes to the network, in terms of concepts, properties and semantic relations. Each Peer interacts with others by submitting discovery queries in order to identify the potential members of an interest-based community, and by replying to incoming queries whether it can join a community. A semantic matchmaker is employed to check whether two Peer share the same interests. We refer the reader to [31] for a brief survey of existing ontology matching approaches. The other drawback of this approach is that a Peer's interests are inevitably revealed, even to the Peers that do not belong to the community; therefore the privacy of the Peer is compromised.

Khambatti et al. proposed a Peer-to-Peer community discovery approach where each Peer is associated with a set of attributes that represent the interests of that Peer [25]. These attributes are chosen from a controlled vocabulary that each Peer agrees with, which gets rid of the uncertainty of the fuzzy ontology matching. Peers whose attributes have non-empty intersection can be grouped together. A very basic privacy policy is applied such that a Peer does not disclose attributes corresponding to its private interests. This means that the smaller the number of claimed attributes, the smaller the number of communities or community members discovered by a Peer. Peer-to-Peer data mining is a relatively new field. It pays careful attention to the distributed resources of data, computing, communication, and human factors in order to use them in a near optimal fashion. To name a few, Wolff et al. proposed algorithms for association rule mining [43] and local l2 norm monitoring over P2P networks [42]. Datta et al. proposed an algorithm for K-Means clustering over large, dynamic networks [13].

## 4. INFORMATION RETRIEVAL IN P2P NETWORKS

Information Retrieval (IR) systems keep large amounts of unstructured or weakly structured data, such as text documents or HTML pages, and offer search functionalities for delivering documents relevant to a query. The main challenge for information retrieval in Peer-to-Peer networks is to be able to guide the query to the other Peers that contain the most relevant answers in a fast and efficient way. The design of scalable models for IR over P2P networks remains an open issue. This motivates us to propose a scalable Peer-to-Peer infrastructure that enables advanced method for query routing and Information Retrieval, and imposes low network and hardware load on the Peers. Today researchers from different areas, including database systems, distributed systems, networking and information retrieval, have started to work on efficient, yet semantically powerful search mechanisms in Peer-to-Peer systems.

Odysseas Papapetrou [46] proposes new approaches for enabling distributed IR over P2P without limiting the network size or mutilating the IR. The basis of these approaches is an innovative distributed clustering algorithm, which can cluster Peers in a P2P network based on their content similarity. This clustering enables significant network savings and enables new families of distributed IR algorithms.

Nottelmann and Fuhr [47] build an IR system over a hierarchical P2P network. The Peers there do not maintain a distributed index; instead, some Super-Peers are assigned the responsibility to keep their Peers' summaries, and to forward the queries to the most related of their Peers, or to other Super-Peers. In addition to the infrastructure, the authors present a decision theoretic model for optimal P2P query routing. For selecting the Peers for each query, their model considers the cost of query routing and the expected results from each Peer. The approach gives expected optimal query results for the query execution cost.

Sharma and al. [48] introduce a system, called IR-Wire, for information retrieval research in the Peer-to-Peer file-sharing domain. This tool maintains many statistics and implements a number of information retrieval ranking functions and contains a data logger and analyzer. The data logger logs both incoming and outgoing queries and query results and provides a way to create a snapshot of the entire data set shared by the users. The data analyzer provides a simple user interface for data analysis. This work was meant to address in the research for tools and data for P2P IR, expressed in [49]. Today's, data management in Peer-to-Peer (P2P) provide a promising approach that offers scalability, adaptively to high dynamics, and failure resilience. Although there exist many P2P data management systems in the literature, most of them focus on providing only information retrieval (IR) [50] [51] or filtering (IF) [52] functionality (also referred to as publish/subscribe or alerting), and have no support for a combined service. DHTrie [53] is an exact IR and IF system that stresses retrieval effectiveness, while MAPS [54] provides approximate IR and IF by relaxing recall guarantees to achieve better scalability.

## 5. SEMANTIC MAPPINGS AND HYPER-GRAPHS

This section is devoted to the study of two methods developed and used for queries routing in P2P communities. The baseline method developed in [19], uses semantic similarity functions to establish semantic mapping between Peers and Peers/Super-Peers. Unfortunately, this approach is not being scale due to the mappings it uses and this problem arise considering only thousands of Peers in the network. This limit motivates our investigation and the development of our new method based respectively on clustring/hypergraphs.

## 5.1 Baseline approach

A new Peer Pj advertises its expertise by sending, to its Super-Peer, a domain advertisement DAj = (PID; $E_{XP}^{j}$, T$_j$ ; $\varepsilon_{acc}$; TTL) containing the Peer ID denoted PID, the suggested expertise $E_{XP}^{j}$, the topic area of interest Tj, the minimum semantic similarity value ($\varepsilon_{acc}$) required to establish semantic mapping between the suggested expertise $E_{XP}^{j}$ and the theme of its Super-Peer. When receiving an expertise $E_{XP}^{j}$, a Super-Peer SP$_a$ invokes the semantic matching process to find mappings between its suggested schema and the received expertise. The results of this procedure are two indexes, i.e. mediation indexes, materializing the semantic similarity between both Super-Peers together and Peers/Super-Peers.

The semantic routing algorithm (Algorithm 1) of baseline approach exploits the expertise of (Super-)Peers and the two levels of mappings in order to forward a query q to only relevant Super-Peers. The algorithms computing the different mappings were introduced in [19]. A Peer $P_2$ submits its query $Q^2$ on its local data schema. This query is sent to his Super-Peer SP$_A$ responsible for the community (See Figure 3). The Super-Peer SP$_A$ in turn suggests, based on the index obtained by the process of mediation (first level), the Peers P1 of his community or the other Super-Peers SP$_P$ that are able to treat this query. Each submitted query received by a Super-Peer, is processed by searching connections (second level of mappings) between the subject of this query and expertise of Peers (of the same community) or the description of themes of other Super-Peers.

In turn, a Super-Peer from the nearby community, having received this request, researches among Peers (in his community) who are able to answer this query. The major problem of this approach is the mediation at the two levels cited above: if we take thousands of Peers or Super-Peers this approach can not be scale due to the mappings at both levels.

---

**Algorithm 1 : Baseline algorithm**
Input: Q : Query
     SP : Super-peer of P
Output: *SR$_Q$* : Set of answers of Q
   1: Variables : PSet : Set of peers
   2: NP : Neighbors of SP (set of super-peers)
   3: *SR$_Q$* = ; 4: PSet = *Capacity$_{CMSP/P}$* (Q) >$\varepsilon_{acc}$
   5: repeat
   6:    *SP$_Q$* = get(*s 2 PSet*);
   7:    Remove *SP$_Q$* from PSet;
   8:    *SR$_Q$* = *SR$_Q$* [ *Query*(*SP$_Q$*);
   9: until (*PSet* = ;))
  10: repeat
  11:    *SP$_Q$* = *Capacity$_{CMSP/SP}$* (Q) > $\varepsilon_{acc}$
  12:    Remove *SP$_Q$* from NP;
  13:    *SR$_Q$* = *SR$_Q$* [ BL(Q; *SP$_Q$*);
  14: until (*PSet* = ;))
  15: Return(*SR$_Q$*);

---

Query routing in these networks is therefore very problematic. Semantic Routing is a method of routing which focuses on the nature of the query to be routed than the network topology. Essentially semantic routing improves on traditional routing by prioritizing nodes which have been previously good at providing information about the types of content referred to by the query. Semantic Routing is obviously not the most optimal solution for routing, and it wasn't long before other P2P routing algorithms emerged which were more efficient.

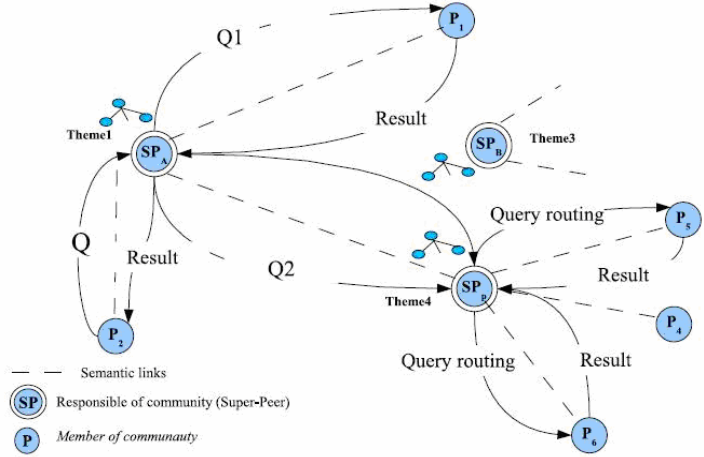

Figure 3. Network configuration and query routing (baseline approach).

Assuming that Peer $P_2$ issues a query $Q^2$, the query routing algorithm proceeds as follows:
- We first find the responsible Super-Peer for $P_1$ which in this example is $SP_A$.
- The responsible Super-Peer ($SP_A$) process the query to find the relevant Peers of his community (ex.: $P_1$) if there are, and also find the others Super-Peers (ex.: $SP_P$) that might content relevant Peers to answer the query.
- Each relevant Super-Peer(s) ($SP_A$, $SP_P$) treat(s) query to find relevant Peers using the function CAP that measures the capacity of a peers of expertise $E_{XP}(P_1)$ on answering a given query of subject of Sub(Q).

$$Cap(P,Q) = \frac{1}{\text{Sub}(Q)} \sum_{S \in Sub(Q)} \underset{e \in Exp(P)}{Max} S_S(s,e)$$

- Then the final set of relevant peers (($P_1:SP_A$)...($P_5:SP_P$)) and their corresponding super-peers are returned. Semantic routing is not a reasonably idea when the network growth. This motivates us to develop a new approach based on clustering super-peer.

The followings sections describe our approach in order to avoid Super-Peer, when it's too busy to treat all users' queries, to process the second level of mapping. This approach improves response times of queries and scalability in P2Ph context by restructuring the network dynamically by introducing the concept of soft clustering to find minimal transversal between clusters.

## 5.2 Knowledge Based Network

A Knowledge Super-Peer (KSP) network is a hybrid semantic sub-network of Overlay Network (HSSON). It combines the semantic network with knowledge to define the HSSON network. In this work, we enrich the P2P network that includes peers and super-peers by the third kind of nodes based on knowledge, i.e. KSP. In fact, we consider a partition of the super-peers space and we associate to each subset of super-peers $C_j$, i.e., the $j^{th}$ community, a node $KSP_j$ defined as a predictive model that return the Super- Peers that may have relevant data to answer queries with minimum query tasks and by consequence, improving answering time of queries. Thus, the node KSPj is represented by a decision tree, denoted TSP, constructed from queries processed successively by it. Details on the construction of this decision tree exceed the scope of this paper. For a more details of the algorithmic aspect of this problem, we refer the reader to [7] [39].

The KSP number j is defined as follows:

$$\text{KSPj} = \bigcup\nolimits_{l=1}^{|M|} (SI_e^l C), M \subseteq T$$

Where M is the number of super-peer in KSP$_j$ and M $\subseteq$ T, T is the total number of super-peers. $SI^l_eC$ is the Semantic Inter-Community of the super-peer number l. Two fundamental properties are derived from KSP:

$$KSP_i \cup KSP_j = SON, \; i \neq j$$

$$KSP_i \cap KSP_j \neq \phi$$

A Knowledge Super-Peer is represented physically with a specific Peer. This Peer, representing the Knowledge Super-Peer number j, is noted as follows:

$$KSP_j = (PS \cup SP_{TJ, DJ}, E_{XP}(PS), K^j, RSC^J, RSI^J, IND^j)$$

where PS$\subseteq$P is a subset of Peers having very close center of interests denoted T J = {T$_1$,..., T$_s$}, EXP (PS) is the set of expertise of Peers interested by at least one of themes in T J, SP$_{T J, DJ}$ (belong to SP) is the set of Super-Peers responsible of communities which have very close domain interests, DJ = {D$_1$, ..., D$_s$} represents the description of themes in T J (DJ describes TJ). Kj $\subseteq$ K is the set of physical links between each Super-Peer S$_{PTj, Dj}$ $\in$ SPT J, DJ and 1. The Peers connected to it (within its community); 2. The other Super-Peers; 3. The Knowledge Super-Peer KSP$_j$ itself. RSC$^J$ is the set of semantic Intra-Community of the Super-Peers $\in$ SP$_{TJ, DJ}$. RSI$^J$ is the set of semantic Inter-Community for each Super-Peer in SP$_{TJ, DJ}$. IND$^j$ is the index obtained using a decision tree algorithm to identify directly the most relevant (Super-)Peers, without going through mappings, to provide good results when a query is submitted by a Peer.

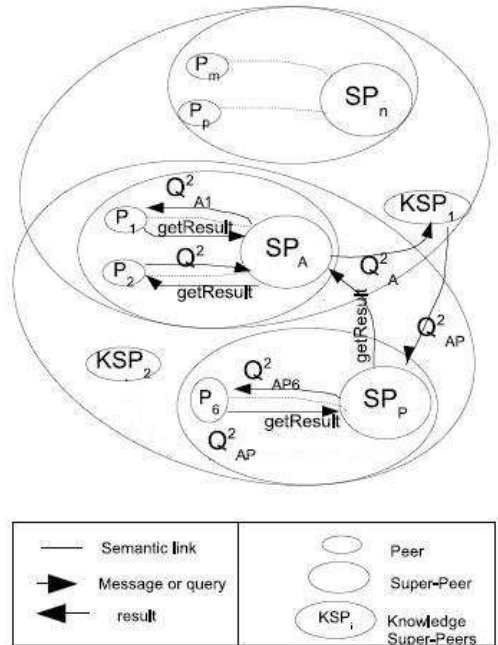

Figure 4: Network configuration and query routing (KSP approach).

Our proposed System (See Figure 4) is a hybrid P2P system based on an organization of Peers around Super-Peers according to their proposed themes, where Super-Peers are connected to a Knowledge-Super-Peer (KSP), the engine that specifies the Super-Peers having Peers which may have relevant data to answer queries with minimum query tasks and, by consequence, improve answering time of the queries. The Super-Peer architecture allows the heterogeneity of Peers by assigning more responsibility to Peers able to assume them. Therefore, certain Peers, called Knowledge-Super-Peers, have an additional computing power and greater bandwidth, resources and performing administrative tasks. They are responsible of routing queries to relevant Super-Peers, allowing not only to reduce efforts of compilation of queries but also to prevent the spread of queries in the network. In each community, there is a Super-Peer connected to a Knowledge-Super-Peer where we have an index to identify Super-Peers that are most relevant to provide good results of queries.

The building block (KSP) of the current P2P systems in the architecture (Distributed Knowledge - DK) is the notion of a Super-Peer-group, or a number of nodes (Super-Peer) that participate with each other for a common purpose to minimize the load in the KSP. The algorithm 2 is used to predict Super-Peers able to answer any submitted query using decision tree. Example : In this example we explain the query routing using KSP (Fig. 4), A Peer $P_2$ sends a query $Q^2$ to his SP ($SP_A$) that in its turn sends this query to KSP that belong to and also to Peers of his community that are able to answer this query. This KSP analyzes the query to find the other SP using decision tree to send this query. Finally, the results will be sent to $P_2$.

**Algorithm 2: Knowledge Based Algorithm**
Input: Q: Query
      SP: Super-peer of P
Output: $SR_Q$: Set of answers of Q
   1: Variables: $T_{SP}$: *decision tree of SP*
   2: NP: Neighbors of SP (set of super-peers)
   3: $SR_Q = \phi$
   4: PSet = Select ($p \in SP$);
   5: repeat
   6: $SP_Q$ = get($s \in PSet$);
   7: Remove $SP_Q$ from PSet;
   8: $SR_Q = SR_Q \cup Query(SP_Q)$;
   9: until ($PSet = \phi$ ))
  10: $SP_Q = T_{SP}(Q)$);
  11: $SR_Q = SR_Q \cup [ Query(SP_Q)$;
  12: Return ($SR_Q$);

### 5.3. Hypergraph Transversals based approach

This section introduces a new efficient method for queries routing in the P2P context that is based on both the Super-Peer clustering algorithm called ECCLAT and the computation of a minimal query routing strategy. The clustering of Super-Peers using their expertise leads to the explicit construction of communities where each one is represented by a set of Super-Peers (cluster of Super-Peers) with the constraint that a Super-Peer may belong to more than one cluster. In this situation the set of clusters constitutes a set of hypergraph and where each node constitutes a community. The question is than how to find the minimal querying strategies where each one is a set of Super-Peers that covers all communities. The function cover means that the minimal set contains at least on Super-Peer of each community. Consequently, strategy guaranties that it represent all expertise of the network. Thus, we consider that a strategy is a semantic context that can be useful for queries routing. In fact, when a Super-Peer SP receive a query Q and can not answer it using only its Peers than it select possible minimal strategy minS where SP$\in$ minS.

A transversal is minimal in the sense that guaranties that all communities (cluster of Super-Peers) are represented:

$$\forall T_c \in T ; \forall c \in C : T_c \cap c \neq \phi ;$$

Where C is the set of communities (Super-Peers clusters), T is the set of transversals.

Table 1. Example of a dataset D1

| Id. | Items | | | | | | | | |
|-----|----|----|----|----|----|----|----|----|----|
| $SP_1$ | $W_1$ | $W_2$ | $W_3$ | | | | | | |
| $SP_2$ | $W_1$ | $W_2$ | $W_3$ | | | | | | |
| $SP_3$ | $W_1$ | $W_2$ | $W_3$ | | | | | | |
| $SP_4$ | | | | $W_4$ | $W_5$ | | | | |
| $SP_5$ | | | | $W_4$ | $W_5$ | | | $W_8$ | |
| $SP_6$ | $W_1$ | | | $W_4$ | $W_5$ | $W_6$ | $W_7$ | $W_8$ | |
| $SP_7$ | $W_1$ | | | | | $W_6$ | $W_7$ | | $W_9$ |
| $SP_8$ | | | | | | | | $W_8$ | $W_9$ |

In our context, we cluster Super-Peers according to their expertise. Table1 presents an example of transactional dataset. There are 8 transactions (denoted $SP_1$… $SP_8$) and 9 items (denoted $W_1$… $W_9$). Transactions correspond to Super-Peers. Items correspond to components of a query successfully processed by the Super-Peers. For example, $W_1$ is present in the transaction $SP_1$ because $W_1$ is a component of a query successfully processed by the Super-Peer SP1. The obtained clusters with minfr=20% and M=1 are: ($W_1$, $W_2$, $W_3$; $SP_1$, $SP_2$, $SP_3$), ($W_4$, $W_5$; $SP_4$, $SP_5$, $SP_6$), ($W_1$, $W_6$, $W_7$; $SP_6$, SP7) and ($W_9$; $SP_7$, $SP_8$).

The cluster ($W_1$, $W_2$, $W_3$; $SP_1$, $SP_2$, $SP_3$) shows that $SP_1$, $SP_2$ and $SP_3$ share an expertise characterized by the association of the components $W_1$, $W_2$ and $W_3$.

Tableau 2. A dataset D2.

| Id. | Items |
|-----|-------|
| $SP_1$ | $W_1$ $W_2$ $W_3$ $W_4$ $W_5$ $W_6$ $W_7$ $W_8$ $W_9$ $W_{10}$ $W_{11}$ $W_{12}$ $W_{13}$ $W_{14}$ $W_{15}$ $W_{16}$ $W_{17}$ $W_{18}$ |
| $SP_2$ | $W_1$ $W_3$ $W_5$ $W_6$ $W_7$ $W_8$ $W_9$ $W_{10}$ $W_{11}$ $W_{12}$ $W_{14}$ $W_{19}$ $W_{20}$ $W_{21}$ $W_{22}$ $W_{23}$ $W_{24}$ |
| $SP_3$ | $W_2$ $W_4$ $W_9$ $W_{18}$ $W_{25}$ $W_{26}$ $W_{27}$ $W_{28}$ $W_{29}$ $W_{30}$ $W_{31}$ $W_{32}$ $W_{33}$ $W_{34}$ |
| $SP_4$ | $W_3$ $W_{17}$ $W_{24}$ $W_{35}$ $W_{36}$ $W_{37}$ $W_{38}$ $W_{39}$ $W_{40}$ $W_{41}$ $W_{42}$ |
| $SP_5$ | $W_2$ $W_4$ $W_{11}$ $W_{12}$ $W_{19}$ $W_{37}$ $W_{40}$ $W_{41}$ $W_{43}$ $W_{44}$ $W_{45}$ $W_{46}$ |
| $SP_6$ | $W_1$ $W_{11}$ $W_{13}$ $W_{14}$ $W_{17}$ $W_{19}$ $W_{24}$ $W_{35}$ $W_{36}$ $W_{37}$ $W_{38}$ $W_{39}$ $W_{40}$ $W_{41}$ $W_{42}$ $W_{45}$ $W_{46}$ $W_{47}$ |
| $SP_7$ | $W_2$ $W_6$ $W_{11}$ $W_{17}$ $W_{20}$ $W_{36}$ $W_{37}$ $W_{38}$ $W_{39}$ $W_{41}$ $W_{42}$ $W_{43}$ $W_{44}$ $W_{48}$ $W_{49}$ $W_{50}$ $W_{51}$ $W_{52}$ $W_{53}$ |
| $SP_8$ | $W_5$ $W_6$ $W_8$ $W_{21}$ $W_{23}$ $W_{24}$ $W_{25}$ $W_{28}$ $W_{30}$ $W_{33}$ $W_{44}$ $W_{54}$ |
| $SP_9$ | $W_6$ $W_{21}$ $W_{36}$ $W_{42}$ $W_{49}$ $W_{50}$ $W_{52}$ $W_{53}$ $W_{55}$ $W_{56}$ $W_{57}$ $W_{58}$ $W_{59}$ $W_{60}$ $W_{61}$ $W_{62}$ $W_{63}$ |
| $SP_{10}$ | $W_{19}$ $W_{37}$ $W_{40}$ $W_{41}$ $W_{45}$ $W_{46}$ $W_{47}$ $W_{64}$ $W_{65}$ |

Table 2 presents another example with 300 Peers and 10 Super-Peers. The resulting clusters minfr=20% and M=1 are:

($W_{19}$, $W_{37}$, $W_{40}$, $W_{41}$, $W_{45}$, $W_{46}$; $SP_5$, $SP_6$, $SP_{10}$)
($W_{17}$, $W_{36}$, $W_{37}$, $W_{38}$, $W_{39}$, $W_{41}$, $W_{42}$; $SP_4$, $SP_6$, $SP_7$)
($W_6$, $W_{21}$; $SP_2$, $SP_8$, $SP_9$)
($W_5$, $W_6$, $W_8$; $SP_1$, $SP_2$, $SP_8$)
($W_2$, $W_4$; $SP_1$, $SP_3$, $SP_5$)

Figure 5 focuses only on the resulted five clusters and an interesting feature of the clustering algorithm used is its ability to produce a clustering with a minimum overlapping between clusters (approximate clustering) or a set of clusters with a slight overlapping. These five clusters are than used to find all minimal transversals of the hypergraph (clusters) to link all the edges (SP) that are belong the traversals route for query routing. The resulted set of transversals is:

Transversals2 = {{$SP_1$, $SP_2$, $SP_6$}, {$SP_1$, $SP_6$, $SP_8$},
{$SP_1$, $SP_6$, $SP_9$}, {$SP_2$, $SP_3$, $SP_6$}, {$SP_2$, $SP_4$, $SP_5$},
{$SP_2$, $SP_5$, $SP_6$}, {$SP_2$, $SP_5$, $SP_7$}, {$SP_3$, $SP_6$, $SP_8$},
{$SP_4$, $SP_5$, $SP_8$}, {$SP_5$, $SP_6$, $SP_8$}, {$SP_5$, $SP_7$, $SP_8$}}
Transversals3 = {{$SP_1$, $SP_2$, $SP_4$, $SP_{10}$},
{$SP_1$, $SP_2$, $SP_7$, $SP_{10}$}, {$SP_1$, $SP_4$, $SP_5$, $SP_9$},
{$SP_1$, $SP_4$, $SP_8$, $SP_{10}$}, {$SP_1$, $SP_4$, $SP_9$, $SP_{10}$},

{SP$_1$, SP$_5$, SP$_7$, SP$_9$}, {SP$_1$, SP$_7$, SP$_8$, SP$_{10}$},
{SP$_1$, SP$_7$, SP$_9$, SP$_1$0}, {SP$_2$, SP$_3$, SP$_4$, SP$_{10}$},
{SP$_2$, SP$_3$, SP$_7$, SP$_{10}$}, {SP$_3$, SP$_4$, SP$_8$, SP$_{10}$},
{SP$_3$, SP$_7$, SP$_8$, SP$_{10}$}}

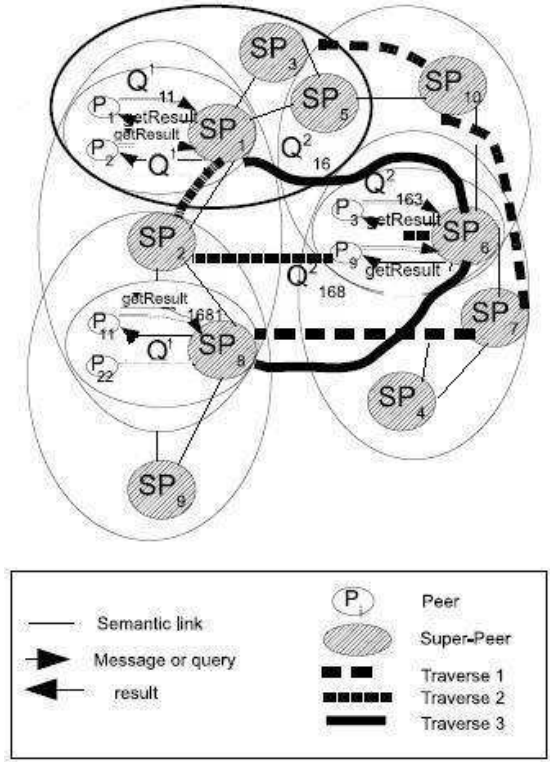

Figure 5. Example of routes in a hypergraph of Super-Peers.

Figure 4 depicts only the three following minimal transversals: {{SP$_1$, SP$_2$, SP$_6$}, {SP$_1$, SP$_6$, SP$_8$} and {SP$_3$, SP$_7$, SP$_8$, SP$_{10}$}}

The following algorithm uses only one minimal traversal (strategy) to answer the query Q asked by the Peer P (algorithm 2):

---

**Algorithm 2 : Use only one strategy (1-Strategy)**

Input: S: set of strategies (minimal transversals)
    Q: Query
    P: the peer that sent the query Q
Output: $R_Q$: An answer of Q
    1: Variables: PS: Set of possible strategies
    2: PS = Select ($s \in 2\ S: P \in s$);
    3: $SP_Q$ = Filter (PS, Q);
    4: $R_Q$ = Query ($SP_Q$);
    5: Return ($R_Q$);

---

The algorithm 2 select only one strategy, set of Super-Peers, and send the query considering only its Super-Peers (belongs to the minimal transversal) then to any relevant Super-Peer while using the function CAP of algorithm 1 to select the most knowledge-able Peer for a giver query. We will consider this algorithm in the next experimental section.

Traverse architecture is a physical redistribution of architecture "Baseline" with groups. Most SP must belong to a cluster at least. All Super-Peers of a cluster are connected together. Therefore, the clusters have at least one Super-Peer in common, used to find the minimum traversals between clusters. The in common Super-Peers are used to route the queries to another clusters. A query sent by a Super-Peer who belongs to a cluster and not belonging to the traversal route, was sent to the Super-Peer that belongs to the traversal route. And consequently towards a Super-Peer(s), of another group, which belongs to the traversal route, then towards the relevant Super-Peers related to this Super-Peer.

Assuming that Peer $P_1$ issues a query $Q_1$, the query routing algorithm proceeds as follows:

- We first find the responsible Super-Peer for $P_1$ which in this example is $SP_1$.

- The responsible Super-Peer $SP_1$ sends the query to the Super-Peer $SP_1$ that belongs to transversal route (transversal route: $SP_1$, $SP_2$, $SP_6$).

- This Super-Peer $SP_1$ will send the query to other Super-Peers $SP_2$, of other cluster, that is on the traversal route, then to the relevant Super-Peer(s) $SP_8$.

- Each relevant Super-Peer treats query to find relevant Peers.

- Then the final set of relevant Peers (($P_2$:$SP_1$), ($P_{11}$:$SP_8$)…) and their corresponding Super-Peers are returned.

## 6. EXPERIMENTS AND EVALUATIONS

We describe the performance evaluation of our routing algorithm with a SimJava-based [55] simulator. All experiments were run on a machine Core 2 Duo 1.83GHZ with 4 GB RAM, 250 GB Hard disk and Windows Vista operating system. In our experimental study we compared the performance of our proposed system (Traverse) with an unstructured system [19] which is always used as the baseline in the evaluation of P2P information retrieval. Evaluating the performance of P2P network is an important part to understand how useful it can be in the real world. As with all P2P applications, the first question is whether P2P is scalable. Our systems were evaluated with different set of parameters i.e. number of Peers, number Super-Peer etc. Evaluation results were quite encouraging. There are many dimensions in which scalability can be evaluated: one important metric is the time it takes the Answer of a given query, precision and recall. We run simulations on P2P network of three different sizes. Each Peer sends Query to its SP that sends the query to the Super-Peer, that belong to the traverse route, in turn it will send the query to other Super-Peer (that also to the traverse route) that is connected to relevant Super-Peer to answer the query.

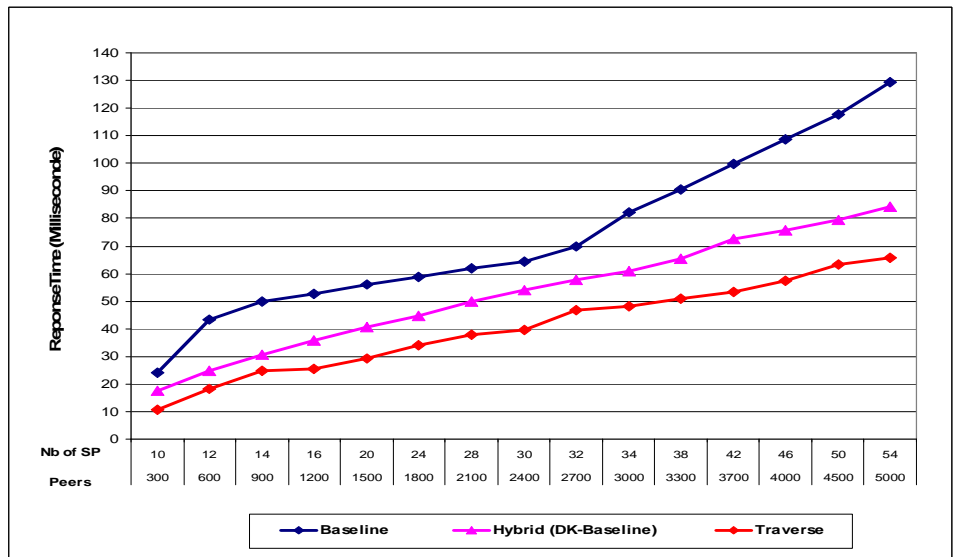

Figure 6. Execution time.

- First one, we modified the number of Peers (300, 600, ..., 5000 Peers) and Super-Peers (10, 12, 14, 16, 20,..., 54) in the both Architectures to measure the execution time and number of messages.

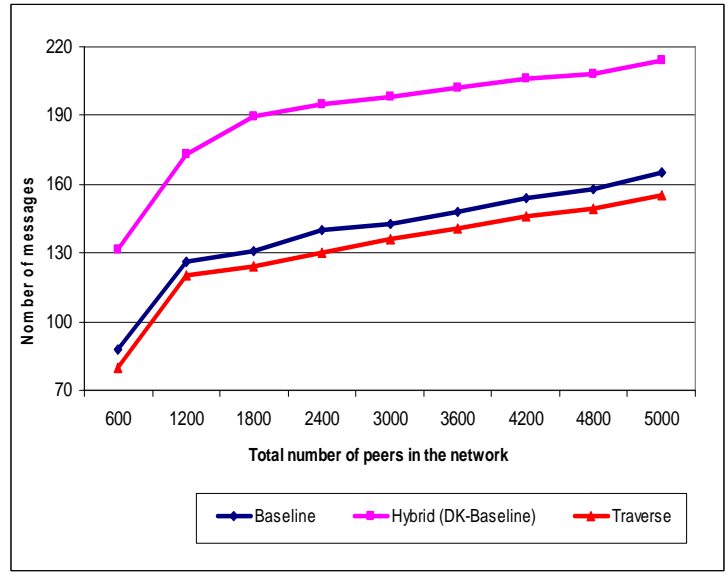

Figure 7: Number of Messages.

- The most popular measure for the effectiveness of our systems is the precision and recall.

$$precision = \frac{\# \text{ of relevant responses retrieved}}{\text{total \# of retrieved Responses}}$$

$$recall = \frac{\# \text{ of relevant responses retrieved}}{\text{total \# of relevant Responses}}$$

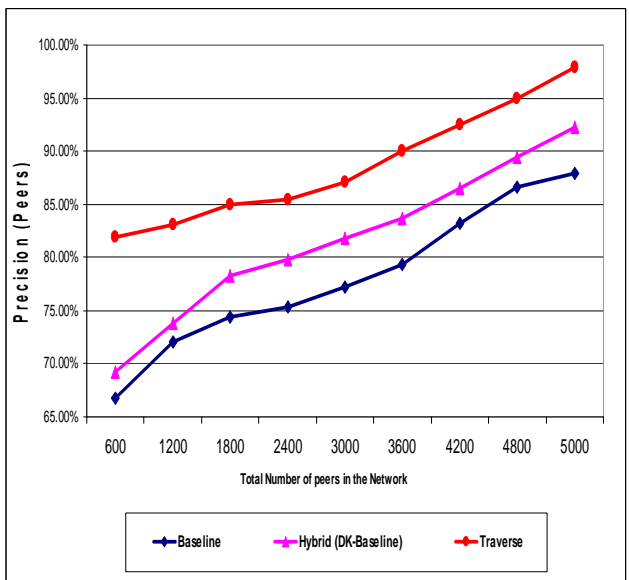

Figure 8: Precision rate

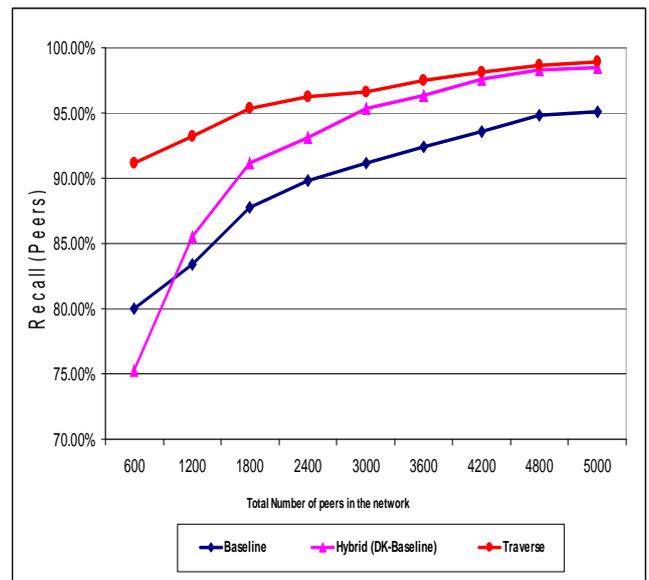

Figure 9: Recall rate

The Graphs shown in figures 6, 7, 8 and 9 are the results of our simulations. They demonstrate the performance of clustering P2P communities (SP) for routing Queries to relevant SP. In the first observation, the difference in the execution times at 300 Peers in the hybrid and in the traverse architectures is small comparing to Baseline architecture (See Figure 6). The execution time was measured as repository size increased. With increasing the numbers of Peers and Super-Peers (more then 600 Peers and 12 Super-Peers), for example at 5000 Peers and 54 Super-Peers, the response time in the hybrid decreases about 35% and in the traverse architecture about 50% comparing to baseline architecture. This means how much our proposed architectures are scalable. Figure 7 shows the variation between the numbers of messages between the Baseline and the hybrid and traverse architectures, where we minimize a little the number of messages in the traverse architecture, this due to the topology of the architectures (baseline and traverse) where we had restructure the baseline architectures to regroup the super-Peers into clusters and use the minimal transversal to route the query, while in the hybrid architecture, The variation between the numbers of messages (See Figure 7) has significant differences, this is due to the presence of a number of KSP (high level of the P2P network) to route the queries to relevant Super-Peers..

Measurements in Figure 8 have shown the precision of the hybrid (81%) and traverse architectures (87%) compared to Baseline architecture (77%). We observe clearly the difference between the proposed architectures (hybrid and traverse) and the baseline architecture, this due to presence of the groups of the Super-Peers that had same similarity of the queries contents and the queries sent to the destination SP, therefore this minimizes the bandwidth consumption of the network which is a problematic of the baseline. This experiment was designed also to measure the accuracy of data which is the recall (See Figure 9).The recall increases with the size of the network and reaches a percentage of almost 95% in hybrid, 96% in the traverse architectures and about 91% in the baseline architecture. These results show the affecting of our mechanisms (using decision tree and clustering of SP) in P2P context, although all architectures are in the nineties concerning the recall. Otherwise, the simulation results show that our mechanism had a remarkable performance in improving the execution time in Peer-to-Peer information retrieval environment. We perform experiments to demonstrate that our proposed system affects performance and improve the scalability of the overall systems.

## 7. CONCLUSIONS

P2P systems are being deployed fairly actively on the Internet. However, the existing systems address different aspects of P2P problems and none of them are perfect. We proposed a Super-Peer topology as a suitable topology for these schema-based P2P networks and discussed how this additional schema information can be used for routing and clustering in such a network. Query routing among Peer communities is based on forwarding policies. We proposed an advanced method using Decision tree to effectively select relevant Peers for a given query and we also used hypergraph based algorithm with minimum traversal to route a given query. The advantage of this model is the robustness in Queries routing and scalability issues in P2P Network with respecting very important issues such as data privacy and the dynamic nature of the underlying network: Peers can leave the overlay network and new Peers can join it. One important area for improvement is performance. Some of the options for improving performance were discussed in the evaluation of P2P Network and include: improvements in the Answering time of a given query and dynamic nature of P2P Network.

The presented time was measured as repository size increased more than 50% in at 5000 Peers traverse architecture less then architecture-baseline. The outcome of these experiments is particularly valuable since it represents the real simulations of our model. The results are in complete agreement with the theoretical predictions and simulations. We believe that there is a need for such flexible query rewriting approach to cope with the high dynamicity and heterogeneity of the Web-based environments. Discovering communities on the fly are essential to perform community directed searching. We show that while our techniques maintain the better quality of results, our techniques reduce response time in P2P search.

We experiment our technique using a Java implementation. By analysis of the outcome of the experiments, we demonstrate that the system indeed shows the scalability and dependability properties predicted by our previous theoretical and simulation results. Since scalability is of great importance in P2P environments, the information space is organized in communities that are inter-related using Peer relationships.

## ACKNOWLEDGEMENTS

This work has been done as a part of the project "recherche intelligente d'information multimedia multilingue arabe" by the franco-libanais comity Cedre.

## REFERENCES

[1] Karl Aberer. (2001) "P-grid: A self-organizing access structure for p2p information systems". CoopIS, pp 179–194.

[2] Karl Aberer and Philippe Cudre-Mauroux and Manfred Hauswirth and Tim Van Pelt, (2004) "Gridvine: Building internet-scale semantic overlay networks". International Semantic Web Conference, pp 107–121.

[3] Reza Akbarinia & Vidal Martins, (2006) "Data management in the appa p2p system". HPDGRID, pp 303-317.

[4] Reza Akbarinia, Esther Pacitti & Patrick Valduriez, (2006) "Reducing network traffic in unstructured p2p systems using top-k queries". Distrib. Parallel Databases journal , pp 67–86.

[5] James Bailey, Thomas Manoukian & Kotagiri Ramamohanarao, (2003) "A Fast Algorithm for Computing Hypergraph Transversals and its Application in Mining Emerging Patterns", ICDM, pp 485–488.

[6] Claude Berge, (1989), "Hypergraphs". North Holland Mathematical Library, Amsterdam.

[7] Kanishka Bhaduri, Ran Wolff, Chris Giannella & Hillol Kargupta, (2008) "Distributed Decision Tree Induction in Peer-to-Peer Systems". In Statistical AnalyNway Thnak yousis and Data Mining Journal, volume 1, pp 85–103.

[8] Carlo Sartiani and Paolo Manghi & Giorgio Ghelli and Giovanni Conforti, (2004) "Xpeer: A self-organizing xml p2p database system", P2PDB, pp 456—465.

[9] Silvana Castano & Stefano Montanelli, (2005) "Semantic self-formation of communities of peers". ESWC, pp 137–151.

[10] Edith Cohen , Amos Fiat & Haim Kaplan (2003) "Associative search in peer to peer networks: Harnessing latent semantics", International Journal of Computer and Telecommunications Networking , Vol. 51, pp 1861-1881.

[11] Arturo Crespo & Hector Garcia-Molina, (2002) "Routing indices for peer-to-peer systems", ICDCS, pp 23.

[12] Isabel F. Cruz, Huiyong Xiao & Feihong Hsu, (2004) "Peer-to-peer semantic integration of xml and rdf data sources", AP2PC, Vol. 3601, pp 108-119.

[13] Souptik Datta, Chris Giannella & Hillol Kargupta, (2006) "K-means clustering over a large, dynamic network" SDM SIAM, pp 153-164.

[14] Mema Roussopoulos, TJ Giuli, Mary Baker, Petros Maniatis, David S. H. Rosenthal, Jeff Mogul, (2005) "P2p or not p2p?", vol. 2, pp 32–43.

[15] Guozhu Dong & Jinyan Li. Mining Border Descriptions of Emerging Patterns from DatasetPairs. Knowledge and Information Systems, vol. 2, pp 178–202.

[16] Nicolas Durand & Bruno Cremilleux. ECCLAT: a New Approach of Clusters Discovery in Categorical Data. Conf. on Knowledge Based Systems and Applied Artificial Intelligence (ES), pp 177–190.

[17] Nicolas Durand, Bruno Crémilleux & Einoshin Suzuki, (2006) "Visualizing Transactional Data with Multiple Clustering for Knowledge Discovery". ISMIS, pp 47–57.

[18] Thomas Eiter & Georg Gottlob (2002) Hypergraph Transversal Computation and Related Problems in Logic and AI. JELIA, pp 549–564.

[19] David Faye, Gilles Nachouki & Patrick Valduriez, (2007) "Semantic query routing in senpeer, a p2p data management system". In NBiS, pages 365–374, 2007.

[20] Gary William Flake, Steve Lawrence & C. Lee Giles (2000) Efficient identification of web communities. ACM SIGKDD, pp 150–160.

[21] Dimitrios Gunopulos, Roni Khardon, Heikkio Mannila & Hannu Toivonen. (1997) "Data Mining, Hypergraph Transversals, and Machine Learning". PODS, pp 209–216.

[22] Peter Haase, (2004) "Bibster: A semantics-based bibliographic peer-to-peer system". ISWC2004, pp 122-136.

[23] Céline Hebert, Alain Bretto, and Bruno Cremilleux. (2007) "A data mining formalization to improve hypergraph transversal computation". Fundamenta Informaticae, IOS Press, Vol. 4, pp 415–433.

[24] Dimitris J. Kavvadias & Elias C. Stavropoulos, (2005) "An Efficient Algorithm for the Transversal Hypergraph Generation", Journal of Graph Algorithms and Applications, vol. 2, pp 239–264.

[25] Mujtaa. Khambatti, Kyung Ryu & Partha Dasgupta, (2002) "Efficient discovery of implicitly formed peer-to-peer communities". International Journal of Parallel and Distributed Systems and Networks, Vol. 4, pp 155–164.

[26] Alexander Loser, Felix Naumann, Wolf Siberski, Wolfgang Nejdl & Uwe Thaden (2003) "Semantic overlay clusters within super-peer networks" DBISP2P, pp 33–47.

[27] Jie Lu & Jamie Callan, (2003) "Content-based retrieval in hybrid peer-to-peer networks". ACM CIKM '03: Proceedings of the twelfth international conference on Information and knowledge management, pp 199–206.

[28] M. T. M. Machine Learning. McGraw-Hill Science/Engineering/Math, March 1997.

[29] Stefano Montanelli and Silvana Castno. (2008) "Semantically routing queries in peer-based systems: The h-link approach". Knowl. Eng. Rev., vol. 1, pp 51–72.

[30] Wolfgang Nejdl, (2002) "Edutella: A p2p networking infrastructure based on rdf", Eleventh International World Wide Web Conference, pp 604 - 615.

[31] Natalya F. Noy, (2004) "Semantic integration: a survey of ontology-based approaches". SIGMOD, vol. 4, pp 65–70.

[32] N. Pasquier, Y. Bastide, R. Taouil, and L. Lakhal. (1999) "Efficient Mining of Association Rules Using Closed Itemset Lattices". Journal if Information Systems, vol. 1, pp 25–46.

[33] Bijan Raahemi, Ahmad Hayajneh & Peter Rabinovitch, (2007) "Peer-to-peer ip traffic classification using decision tree and ip layer attributes", International Journal of Business Data Communications and Networking, vol. 4, pp 60–74.

[34] Sylvia Paul Ratnasamy and Sylvia Paul Ratnasamy and Sylvia Paul Ratnasamy, (2001). "A scalable content-addressable network". ACM-SIGCOMM, pp 161–172.

[35]     Antony Rowstron , Peter Druschel, (2001) "Pastry: Scalable, decentralized object location and routing for large-scale peer-to-peer systems", In Lecture Notes in Computer Science book, pp 329–350.

[36]     Ion Stoica, Robert Morris, David Karger, Frans Kaashoek & Hari Balakrishnan (2001) "Chord:A scalable peer-to-peer lookup service for internet applications", ACM SIGCOMM, pp 149–160.

[37]     Christoph Tempich , Steffen Staab , Adrian Wranik. (2003) "Peer-to-peer information retrieval using self-organizing semantic overlay networks", ACM-SIGCOMM, pp 175-186.

[38]     Christoph Tempich, Steffen Staab & Adrian Wranik, (2004) "Remindin: Semantic query routing in peer-to-peer networks based on social metaphors". In Proceedings of the 13th International World Wide Web Conference, pp 640–649.

[39]     Emily Thomas, (2004) "Data mining: Definitions and decision tree examples". In he Association for Institutional Research and Planing Officers(AIRPO).

[40]     Xi Tong, Dalu Zhang & Zhe Yang, (2005) "Efficient content location based on interest-cluster in peer-to-peer system" , ICEBE, pp 324–331.

[41]     Patrick Valduriez & Esther Pacitti, (2005) "Data management in large-scale p2p systems", VECPAR 2004, pp 104–118.

[42]     Ran Wolff and Kanishka Bhaduri & Hillol Kargupta, (2006) "Local l2-thresholding based data mining in peer-to-peer systems", SDM SIAM, pp 430-441.

[43]     Ran Wolff & Assaf Schuster, (2003) "Association rule mining in peer-to-peer systems", ICDM Proceedings of the Third IEEE International Conference on Data Mining, pp 363.

[44]     Beverly Yang & Hector Garcia-Molina, (2002) "Improving search in peer-to-peer networks". In ICDCS '02: Proceedings of the 22nd International Conference on Distributed Computing Systems (ICDCS'02), pp 5.

[45]     Zhou Zhu & James Bailey, (2006) "Fast Discovery of Interesting Collections of Web Services", In Proc. International Conference on Web Intelligence (WI'06), pp. 152–158

[46]     Odysseas Papapetrou, (2008) "Full-text indexing and Information Retrieval in P2P Systems", ACM International Conference Proceeding Series, EDBT, Vol. 326, pp 49-57.

[47]     Henrik Nottelmann & Norbert Fuhr, (2007) A decision-theoretic model for decentralised query routing in hierarchical peer-to-peer networks. In ECIR, pp 148-159.

[48]     Shefali Sharma, Linh Thai Nguyen & Dongmei Jia, (2006),"IR-Wire: A Research Tool for P2P Information Retrieval" ACM SIGIR. pp 6–11.

[49]     Henrik Nottelmann, Karl Aberer, Jamie Callan & Wolfgang Nejdl, (2005), CIKM P2PIR Workshop Report, http://p2pir.is.informatik.uni- duisburg.de/2005/report.pdf

[50]     Sebastian Michel, Matthias Bender, Nikos Ntarmos, Peter Triantafillou, Gerhard Weikum & Christian Zimmer (2005) "MINERVA: Collaborative P2P Search". VLDB, pp 1263-1266.

[51]     Jie Lu & Jamie Callan, (2005) "Federated Search of Text-based Digital Libraries in Hierarchical Peer-to-Peer Networks". In ECIR, pp 477 – 498.

[52]     Ioannis Aekaterinidis & Peter Triantafillou, (2006) "PastryStrings: A Comprehensive Content-Based Publish/Subscribe DHT Network". ICDCS, pp 23.

[53]     Christos Tryfonopoulos, Stratos Idreos & Manolis Koubarakis, (2005) "Publish/Subscribe Functionality in IR Environments using Structured Overlay Networks". ACM-SIGIR, pp 322–329.

[54]     Christian Zimmer, Christos Tryfonopoulos, Klaus Berberich, Manolis Koubarakis & Gerhard Weikum, (2008) "Approximate Information Filtering in Peer-to-Peer Networks". WISE, vol. 5175, pp 6-19.

[55] Fred Howell & Ross McNab, (1998) "simjava: a discrete event simulation package for Java with applications in computer systems modelling", International Conference on Web-based Modelling and Simulation, pp 51-56.

**Authors**

Anis Ismail, Born in Lebanon, April 1979, work as system and network administrator and instructor at Lebanese University, University institute of technology, saida, Lebanon. He received the B.S. degree in Telecommunication and Networking Engineering from Lebanese University (LU), the M.S. in computer science from the American University of Science and Technology (AUST) in Lebanon. Third year PHD at LSIS, AIX Marseille, France.

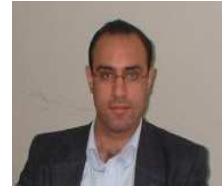

Mohamed Quafafou did his PhD Thesis in 1992 on Intelligent Tutoring Systems at INSA de Lyon, France. From 1992 to 1994, he was ATER at INSA de Lyon and than at Nantes Faculty of Sciences. From 1995 to 2001, he was assistant professor at the Nantes University. During that period, he developed research on Rough Set Theory, concepts approximation, data mining, web information extraction and participated actively with France Telecom to the project Comminges to design a new web system dedicated to French web analysis for discovering emergent web communities. He was also chief-scientist at GEOBS where he headed the Geobs Data Analyzer project, which was developing a spatial data mining systems with application to environment, marketing, social analysis, etc. From September 2002, he was professor at the Avignon University and moved in 2005 to the Aix-Marseille University where he joined the Information and System Science Laboratory (UMR CNRS 6168) and continue his research on web data mining considering different application contexts like P2P, multimedia and web services. Since 2002, he teaches foundations of data/knowledge based systems including machine learning, data mining, personalization, datawarehousing, XML, web services, multimedia, web and mobile applications.

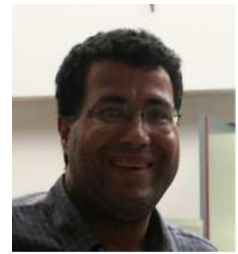

Nicolas Durand, born in France, November 1977, is an assistant professor in computer science at the Aix-Marseille University, Marseille, France. He received the M.S. degree in computer science, in 2000 and the Ph.D. degree in computer science from the University of Basse-Normandie, Caen, France, in 2004. His main research interest covers data mining (clustering algorithms, pattern discovery, hypergraphs) and the applications in P2P networks, Web services, and multimedia.

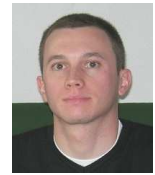

Gilles Nachouki is assistant professor at Nantes University in France. He received a PhD in Computer Science from the University of Toulouse. His interest domain concerns distributed databases. Currently his principal works lie in the domain of data management in peer-to- peer systems.

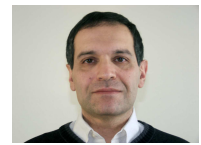

Mohammad Hajjar is a Professeur at University Institute of Technology, Lebanese University, in Lebanon. He received a PHD in computer Science at Nantes University in France. His Interest domain concerns Arabic language processing, multimedia information research and data management in peer-to- peer systems.

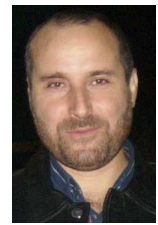